\documentclass[structabstract]{aa}  
\usepackage{graphicx}
\usepackage{natbib}
\usepackage{txfonts}
\newcommand \hi  {\ion{H}{i}}

\newcommand{\kms}{km\,s$^{-1}$}

\newcommand{\ra}{$\alpha$}
\newcommand{\dec}{$\delta$}
\newcommand{\VLSR}{$\upsilon_{lsr}$}
\newcommand{\unith}{10$^{18}$ cm$^{-2}$}
\newcommand{\FWHM}{$ \Delta \upsilon_{1/2}$}
\newcommand \NH  {$N_{\rm HI}$}
\newcommand \msun{M$_{\sun}$}
\newcommand{\lfd}{\hbox{$^{{\rm d}}$}}
\newcommand{\lfh}{\hbox{$^{{\rm h}}$}}
\newcommand{\lfm}{\hbox{$^{{\rm m}}$}}
\newcommand{\lfs}{\hbox{$\mkern-4mu^{\prime\prime}$}}

\newcommand{\lfarcm}{\hbox{$^\prime$}}
\newcommand{\lfarcs}{\hbox{$^{\rm s}$}}
\newcommand{\TKIN}{$T_{kin}$}
\newcommand{\rms}{$\sigma_{n}$}
\newcommand{\TB} {$T_{B}$ }

\begin{document}

\title{Properties of extra-planar \hi~ clouds in the outer part of the Milky
  Way.}  \subtitle{}

 \author{L. Dedes \inst{1,2} \and  P.W.M Kalberla \inst{1}}

 \institute{Argelander Institut f\"ur Astronomie, Universit\"at Bonn,
   Auf dem H\"ugel 71, 53121 Bonn \and
   Max-Planck-Institut f\"ur Radiostronomie, Auf dem H\"ugel 69, 53121 Bonn\\
   \email{ldedes@astro.uni-bonn.de, pkalberla@astro.uni-bonn.de} }

   \authorrunning{L. Dedes\  \&   P.\,M.\,W. Kalberla } 

   \titlerunning{Extra-planar \hi~ clouds}

  \offprints{L. Dedes}

   \date{Received ------; accepted --------}

% \abstract{}{}{}{}{} 
% 5 {} token are mandatory
 
  \abstract
%  {The properties of \hi~ clouds in the disk-halo interface of the Milky Way are not well known.}
%aims
  {There is mounting evidence for an extra-planar gas layer around the Milky
    Way disk, similar to the anomalous \hi~ gas detected in a few other
    galaxies. As much as 10\% of the gas may be in this phase.}  {We analyze
    \hi~ clouds located in the disk-halo interface outside the solar circle to
    probe the properties of the extra-planar \hi~ gas, which is following Galactic rotation.}
%methods
  {We use the Leiden/Argentine/Bonn (LAB) 21-cm line survey to search for \hi~
    clouds which take part in the rotation of the Galactic plane, but are
    located above the disk layer. Selected regions are mapped with the
    Effelsberg 100-m telescope. Two of the \hi~ halo clouds are studied in
    detail for their small scale structure using the Westerbork Synthesis
    Radio Telescope (WSRT)\thanks{The Westerbork Synthesis Radio Telescope is
      operated by ASTRON (Netherlands Foundation for Research in
      Astronomy) with support from the Netherlands Foundation for Scientific
      Research NWO.} and the NRAO Very Large Array (VLA)\thanks{The National
      Radio Astronomy Observatory is a facility of the National Science
      Foundation operated under cooperative agreement by Associated
      Universities, Inc. }.}
%results
  {Data from the 100m telescope allow for the parameterization of 25 distinct \hi~ halo
    clouds at Galactocentric radii 10 kpc $<R<$15 kpc and heights 1 kpc$ <z<$
    5 kpc. The clouds have a median temperature of 620 K,  column
    densities of \NH$\sim10^{19}$cm$^{-2}$, and most of them are surrounded by an
    extended envelope of warmer \hi~ gas. Interferometer observations for two
    selected regions resolve the \hi~ clouds into several arc-minute sized
    cores. These cores show narrow line widths (\FWHM$ \sim 3$ \kms), they
    have volume densities of $n>1.3$ cm$^{-3}$, masses up to 24 \msun, and are
    on average in pressure equilibrium with the surrounding
    envelopes. Pressures and densities fall within the expectations from
    theoretical phase diagrams ($P$ vs $\langle n_{h} \rangle$). The \hi~
    cores tend to be unstable if one assumes a thermally bistable medium, but
    are in better agreement with models that predict thermal fragmentation
    driven by a turbulent flow.  } {}
%{The extra-planar \hi~ gas clumps  detected previously in the inner parts of theMilky Way appears to be wide-spread. }

\keywords{Galaxy: halo - radio lines: ISM - ISM: clouds}
\maketitle
%
%________________________________________________________________

\section{Introduction}
\label{sec_1}

 The \hi~ gas is a major constituent of the interstellar medium (ISM), and it is
a well known property of this gas that it settles in the Galactic plane. From
the very first \hi~ observations it is also known that the disk gas is
co-rotating with the stellar disk. Taking both properties together one may use
the gas distribution to describe the morphology of the Galactic disk.
   
As yet, there is no sharp boundary for the disk emission. \citet{Oort1962dm1}
was the first who mentioned this fact.  ``Well outside the real disk one still
finds neutral hydrogen with an average density of between 5 and 10 per cent of
the intensities one observes in the plane''. Oort was referring to
observations with the Dwingeloo telescope, and \citet{Shane1967IAUS} described
this gas later as a ``galactic envelope'', a smooth envelope of neutral
hydrogen surrounding the spiral structure, following the same Galactic
rotation as the gas in the plane. Further discussion of this envelope was
given by \citet{Takakubo1967BAN} and by \citet{Shane1971AAS}, but there was
some concern about a possible contamination by stray radiation from the
antenna diagram of the Dwingeloo telescope.

The extra-planar gas component was also visible in the \citet{WW1973AAS}
survey, and \citet{Lockman1984} studied this feature in some more
detail. Supplementing observations were made with the NRAO 300-foot telescope
and came also from the NRAO 140-foot survey by \citet{Burton1983AAS}. Lockman
argued that his analysis was not affected by stray radiation. He found for
Galactocentric radii $4 \la R \la 8$ kpc that 13\% of the \hi~ gas is located
outside the disk, extending to $z$-distances of 1 kpc or more and termed this
component an ``\hi~ halo''.

The Bell Labs survey \citep{Stark1992ApJS} is only little affected by
instrumental effects, and \citet{Lockman1991} used this survey to analyze the
nature of the vertical \hi~ gas distribution in the direction of the Galactic
poles. They proposed for the \hi~ gas a decomposition in several layered
structures, corresponding to distinct different isothermal cloud
populations. The scale height for each component results from the pressure
balance of a cloudy turbulent medium against the gravitational potential of
the Milky Way. The concept of a layered structure of the \hi~ contains
essentially three components: a cold neutral medium (CNM), a warm neutral
medium (WNM) and an extra-planar component \citep{Dickey1990ARAAD} which
is often called ``Lockman Layer''. 

The layer concept is based on the average emission from the extra-planar \hi~
gas layer which is very faint. The clumpy nature of the \hi~ gas implies then 
that extra-planar \hi~ clouds must have a low volume filling factor. The layer
concept describes therefore an ensemble of \hi~ clouds or the probability
distribution of such objects. First indications for a population of such
clumps were found by \citet{Simonson1971AA}. These early
data came from the Dwingeloo telescope. Almost three decades later the
Leiden/Dwingeloo survey (LDS,\citep{Hartmann1997}) became available and
provided a much improved database, more sensitive and essentially free of
stray radiation. Channel maps show numerous clumps and filaments that are
detached from the disk, and \citet{KalberlaWestphalen1998} argued for an
extra-planar gas layer which can be characterized by a distribution with a
velocity dispersion of $\sigma = 60$ \kms, considerably larger than the
dispersion suggested by \citet{Lockman1991}.

The first high resolution data of the extra-planar gas layer at a beam-width
of 9\arcmin~ have been taken by \citet{Lockman2002} with the Robert C. Byrd
Green Bank Telescope (GBT). These observations demonstrated convincingly the
nature of the extra-planar gas layer as a population of cold clumps with a
typical mass of 50 \msun. Many of these clumps appear to be surrounded by
warmer envelopes. A larger sample of clouds in the lower halo was
studied by \citet{Ford2008ApJ} with the Parkes Telescope. These clumps are
somewhat larger and more massive than the sample detected by
\citet{Lockman2002}. This cloud population, located close to $R \sim 3.8 $ kpc,
is interpreted as originating from a Galactic fountain. Such a model would
also explain the high kinetic energy which is needed for individual clouds to
reach large $z$-distances. Support for such an interpretation comes from
\citet{Stil2006}. They found fast moving clumps in the Galactic plane with
velocity vectors located {\it within} the Galactic plane, analogous to fast
velocities {\it perpendicular} to the plane as suggested as an explanation for
the extra-planar gas layer.

So far we discussed predominantly Galactocentric distances $R \la 8.5$ kpc,
where 8.5 kpc is the I.A.U Sun--Galactic center distance, since most of the observations are in this range. For a more general description of
this phenomenon, in particular for the question whether the extra-planar gas
layer is caused by a fountain flow, objects at larger distances are
needed. \citet{Kalberla2008AA} argue that extra-planar gas is present even
at $R \ga 35$ kpc. Gas at such distances can hardly originate from fountain
events.

Direct evidence for a population of extra-planar \hi~ clouds outside the
Solar circle was first given by \citet{Snezana2006}. Arecibo data in
the direction towards the anti-center suggest that these clouds are not restricted to
the inner part of the Milky Way disk, which is similar to preliminary results with the
Effelsberg telescope reported by \citet{Kalberla2005ASPC}. In the following we
intend to explore the extra-planar gas layer of the outer part of the Milky
Way in some more detail. Our results are based on single-dish observations
with the Effelsberg 100-m radio telescope and on interferometer observations
with the VLA and the WSRT array. 

This paper is organized as follows. In Sect. \ref{sec_2} we explain our
selection criteria for targets that have been mapped with the 100-m
telescope. Our observations are described in Sect. \ref{sec_3}. We discuss the
properties of the \hi~ cloud sample detected by us with the 100-m telescope,
and also the results from two targeted interferometer observations in
Sect. \ref{sec_4}. We find evidence for a multi-phase structure and compare
in Sect. \ref{sec_6} the derived physical parameters with theoretical
models. Sect. \ref{sec_7} gives our summary and conclusion.

\section{Methods}
\label{sec_2}

Extra-planar \hi~ clouds are known to have a patchy distribution. Accordingly,
a strategy is needed for a successful search. If one likes to measure ``gas
well outside the real disk'' \citep{Oort1962dm1}, the first step is obviously
to determine the extension of the disk. Next, one needs to search in regions
above the disk. For the inner part of the Galaxy the approach is easy. The
scale height of the gas is approximately constant for $ 3 \la R \la 8$ kpc,
the boundary between disk and halo is well defined.  13\% of the gas resides
at $|z| > 500$ pc with little fluctuations for $ 4 \la R \la 8$ kpc
\citep{Lockman1984}.

For the outer part of the Milky Way the situation is more complex. The gas
flares strongly, and in addition the disk is significantly warped. Both cannot be
disregarded, and it is necessary to obtain good estimates for the mid-plane
position and the scale height of the \hi~ gas. Two groups have recently
independently determined the shape of the \hi~ gas distribution, \citet{Levine2006Warp}
and \citet{Kalberla2007DM}. Their results are in good agreement, and we adopt
the disk parameters as derived by \citet{Kalberla2008AA}, which were previously
also used by \citet{Kalberla2007DM} for a determination of the average
extra-planar gas fraction. On average 10\% of the \hi~ gas is located outside
the disk, the extra-planar gas is well defined for $8.5 \la R \la 22$ kpc, but 
tends to increase toward smaller radii $R$, consistent with the determination
by \citet{Lockman1984}.  We conclude that a search for extra-planar \hi~
clouds in the outer part of the Milky Way should be promising for $ R \la 22$
kpc if warp and flaring are taken into account.

\subsection{\hi~ halo cloud selection}
\label{sec_2a}
  
We used the \citet{Kalberla2008AA} model to calculate the expected emission
$t_{ex}(l,b,v)$ for extra-planar gas and $t_{disk}(l,b,v)$ respectively  for
disk gas. The ratio $t_{ex}(l,b,v)/t_{disk}(l,b,v)$ defines a probability that
a cloud feature, observed at position $l,b$ with the velocity $v$ may belong
to the extra-planar gas layer. The extra-planar gas layer is patchy, and  we
therefore searched the Leiden/Argentine/Bonn (LAB) survey
\citep{Kalberla2005ASPC} for positions containing weak \hi~ emission that may
originate from clouds in the lower halo.

We tested our method in the inner part of the Milky Way and recovered those
regions that have been observed previously by \citet{Lockman2002} and
\citet{Ford2008ApJ} as the ones that are most promising for a detection of
extra-planar gas clumps. After this successful test we applied our search
algorithm to $R \ga 8.5$ kpc.

\subsection{Distance determination}
\label{sec_2b}

Quantities directly observable for \hi~ clouds are: column density \NH,
angular size of the cloud $s$ and line width \FWHM~. The cloud diameter $D$, the average spatial
volume density $\langle n \rangle$, pressure $P$ and visible mass $M_{\hi}$ can be
determined only if the distance $d$ of the cloud is known. Since the regions
probed by us are outside the solar circle, a Milky Way velocity field needs to
be used to convert the line-of-sight velocity \VLSR~ of a cloud to its
distance $d$. We use a mass model and a rotation curve according to
\citet{Kalberla2007DM}, which assumes that the halo gas is slightly lagging
behind the Galactic disk. Assuming co-rotation would lead to deviations of
6--22\% depending on the region.  For the \citet{Brand1993} rotation curve,
differences would amount to 15--30\% . Finally, in comparison with the Milky
Way model from \citet{Gomez2006} the deviation for the distance determination
is between -7\% and -14\%. All together, distances determined by us may 
have typical systematical uncertainties of about 15--20\%.

\section{Observations}
\label{sec_3}
\subsection{Effelsberg observations}
\label{sec_3a}

22 fields, each covering 3\degr x3\degr, could be observed with the 100-m
Effelsberg telescope. All the fields have a longitude of $l > 90$\degr, and the
selection was based on the criteria described in Sect. \ref{sec_2}.
Therefore we are confident that the \hi~ emission is associated with the
neutral component of the gaseous halo. The observations were done during the
period from May 2004 to October 2005. We used the AK-90 auto-correlator with
two polarizations at a bandwidth of 10 MHz with 2048 channels. This results in
a channel separation of 4 kHz (velocity separation 0.84 \kms) and velocity
resolution 1.03 km/s. The fields were
mapped beam-by-beam on a 9\arcmin~ grid. The integration for each position was
60 sec. For a system temperature $T_{sys}=27K$ this implies a sensitivity of
\rms=0.1 K. The Effelsberg data were calibrated using the IAU standard
position S7 (Kalberla et al. 1982) . A first order polynomial was applied to
correct the baseline, and the stray radiation contamination was removed using
the method by \citet{Kalberla1980}. The final result was an image cube with a
9\arcmin~ angular resolution and 1 \kms~ velocity resolution. To verify the
observations, the detected \hi~ clouds were re-observed using a full sampling
(4.5\arcmin~ grid). The configuration of the AK-90 auto-correlator was
identical. The integration was increased to  90sec per position, resulting in
a theoretical sensitivity of \rms=0.08 K.

\begin{figure}
\centerline{
\includegraphics[clip,trim=60 30 70 0,scale=0.4]{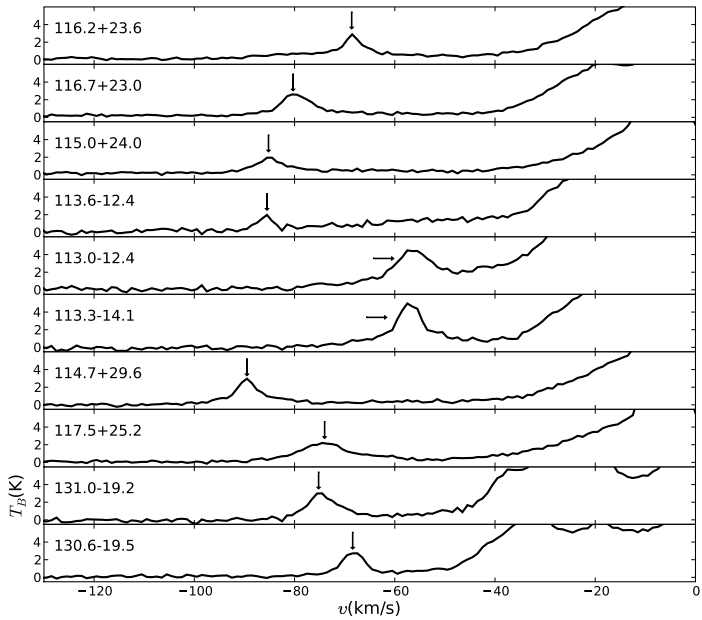}}
\caption[]{Spectra  of \hi~ clouds detected using the Effelsberg
  telescope. Arrows mark the position of the \hi~clouds in the spectra.
The main galactic line is visible at the right side of the
    plot. The extending wings are seen up to $\sim$-100\kms, underlying the \hi~
clouds.}
\label{12673fig1} 
\end{figure}
%trim=60 30 70 60,

\subsection{Synthesis array observations}
\label{sub_3b}

 After extracting a sample of the \hi~ halo clouds from the Effelsberg data
(see Sect. \ref{sec_4a}), follow-up observations were made for two of the
clouds with the WSRT and the VLA synthesis arrays.

The cloud at l,b=116.2\degr,23.6\degr~ at \VLSR=-68 \kms~ was observed with
the WSRT array, mapping the region at (J2000) \ra,\dec=20\lfh4\lfm22\lfs,
82\lfd56\lfarcm36\lfarcs~ in a maxi-short configuration. The integration time
was 12 hours. A double IF was used with a bandwidth of 2.5 MHz and 1024
channels. This backend configuration results in a channel separation of 2.5
kHz and a velocity separation of 0.5 \kms. The data were reduced with the
MIRIAD\footnote{http://www.atnf.csiro.au/computing/software/miriad/} software
package. For flux calibration the source 3C286 was used, while self-calibration
was applied to correct the phase errors. The dirty cube has a sensitivity of
\rms $\sim$2.25 mJy/beam. After applying continuum subtraction, the dirty cube
was de-convolved using the Clark CLEAN algorithm \citep{Clark1980} and
convolved with a Gaussian beam. The final result is an image cube with a
60\arcsec~ resolution and 1 \kms~ velocity resolution.

The cloud at l,b=115.0\degr,+23.9\degr~ and \VLSR=-84.50\kms~ was observed with
the VLA array mapping the region (J2000) (\ra,\dec)=20\lfh 29 \lfm , 82 \lfd
08 \lfarcm~ for six hours in the DnC configuration. A double IF was used with a
bandwidth of 0.78 MHz and 256 channels in each IF. With this configuration we
have a channel separation of 3.05 KHz resulting in a velocity separation of
0.64 \kms. The data were reduced using the NRAO Astronomical Image Processing
System (AIPS)\footnote{http://www.aips.nrao.edu/aips\_faq.html}. The source
3C286 was used as a flux calibrator and the close-by source 2344+824 as a
phase calibrator. The dirty cube has a sensitivity of \rms~ $\sim$2.25
mJy/beam.  The dirty image was de-convolved using the Clark CLEAN algorithm
\citep{Clark1980} after the continuum subtraction. The clean components were
restored with a Gaussian beam, resulting in a clean image with a 60\arcsec~
resolution.

\section{Results}
\label{sec_4}

\subsection{Single dish}
\label{sec_4a}

%TABLE
\begin{table*}
\caption[]{Properties of the observed \hi~ halo clouds.}
\begin{center}
\begin{tabular}{c|| c c c c c c c c c c c}
\hline
            & \VLSR & $R$ & $z$ & $N_{HI}$ & $s$ &$D$ &\FWHM & $T_{kin}$&     $\langle n \rangle$           &   $P$    &  $M_{\hi}$  \\
             & \kms&kpc &kpc     & \unith &arcmin&pc & \kms   &K       & $\mathrm{cm^{-3}}$ &  $K\cdot\mathrm{cm^{-3}}$ &     \msun    \\
\hline
\hline
113.3+27.0  & -40 &10.5 & 1.8 & 16(2) & 13&16(7)& 7.0  & 1080(150) & 0.33(0.15) & 360(170)& 25(22)    \\
113.2+25.5  & -32 &10.0 & 1.4 &  6(1) &  9&8(5) & 4.0  &350(90)  & 0.23(0.14) &  80(50) &  3(4)   \\
116.2+23.6  & -68 &12.5 & 2.8 & 11(2) & 25&51(17)&3.0  &200(70)  & 0.07(0.03) &  14(8)  & 120(80) \\
116.7+22.8  & -81 &14.5 & 3.5 & 25(3) & 27&70(19)&5.1  &570(110) & 0.12(0.04) &  68(26) & 770(440) \\     
116.5+21.4  & -42 &10.5 & 1.5 & 36(4) & 13&14(6) &5.4  &650(120) & 0.70(0.30) & 455(210)&  45(39) \\
115.0+23.9  & -84 &14.5 & 3.9 & 15(3) & 22&60(20)&3.3  & 240(70)  & 0.08(0.03) &  19(7) & 220(140)  \\
115.4+22.4  & -66 &12.0 & 2.5 & 35(3) & 18&34(12)&10.0  &2200(220) & 0.33(0.12) & 730(275)& 260(180)    \\ 
117.3+24.0  & -71 &13.0 & 3.0 & 36(3) & 27&58(18)&8.1  &1440(170) & 0.20(0.06) & 290(90) & 770(490)    \\
118.0+24.6  & -71 &13.0 & 3.1 & 20(3) & 18&40(15)&5.0  &550(110) & 0.16(0.07) &  87(40) & 200(150) \\
118.0+24.0  & -72 &13.0 & 3.0 & 74(4) & 18&40(15)&21.1  &9800(460) & 0.60(0.22) &5800(2140)&750(550)  \\
117.5+25.2  & -74 &13.0 & 3.0 & 60(4) & 18&40(15)& 11.6 &2960(260) & 0.48(0.18) &1449(550) &600(440)  \\
114.5-15.9   & -38 &10.5 &-1.0 &10(2) & 18&18(7) & 4 &350(90)  & 0.18(0.08) &  62(31) &  20(16) \\
113.6-12.4  & -85 &4.5 &-1.9 &  16(2) & 13&35(14)& 4 &350(90)  & 0.14(0.06) &  50(25) & 120(100) \\
113.0-12.4  & -55 &12.0 &-1.2 & 41(3) & 18&28(9) & 6.7 &990(150) & 0.49(0.16) & 470(170)& 205(130) \\ 
113.0-13.0  & -40 &10.5 &-0.9 &  28(3)&  9&9(5) & 4.4  &425(100) & 1.02(0.32) & 435(170)& 14(16) \\
113.3-14.1  & -57 &12.0 &-1.3 & 39(4) &   &&   4.9     &510(100) &            &  &  \\
113.6-13.5  & -50 &11.5 &-1.3 & 68(5) &   &&   7.0     &1080(150) &            &  &    \\ 
114.3+21.8 & -75 &13.5 & 3.0 &  7(-)  &  9&21(12)&    &       &            &  &   \\
128.8-18.5  & -58 &14.0 &-1.9 & 54(6) &   &   & 4.8    &510(100) &            &  &  \\
131.0-19.2  & -72 &15.0 &-2.8 & 49(5) &   &  &  5.4    &640(120) &            &  &  \\
130.6-19.5  & -68 &14.0 &-2.5 & 35(3) &   &  &  6.0    &780(130) &            &  &  \\
112.1+27.3  & -39 &10.5 & 1.8 &  6(1) &  9&10(6) &5.3   &620(120) & 0.20(0.12) & 126(80) & 4(5)  \\
113.8+28.7  & -69 &12.5 & 3.6 &  9(2) & 18&39(14)&6.0  &790(130) & 0.08(0.03) &  61(25) & 90(70) \\ 
112.4+29.9  & -65 &12.5 & 3.7 &  4(1) &  9&20(11)&2.0  &80(40)  & 0.07(0.04) &   6(5) & 11(13)  \\
114.7+29.6  & -89 &15.0 & 5.4 & 19(3) &   &      & 3.6 &285(80)  &            &    &    \\ 
\hline
\end{tabular}
\label{table1}
\end{center}
Note - \VLSR~ is the line of sight
  velocity in \kms, $R$~ is the Galactocentric distance. $z$~ is the height
  above mid-plane. $N_{HI}$ is the \hi~column density. $s$ is the measured
  angular diameter. $D$~ is the spatial diameter. \FWHM~ is the full width half
  maximum of the line width. $T_{kin}$~ the kinetic  temperature. $\langle n \rangle$~ the volume \hi~ density. $P$ is the pressure of the \hi~ gas. $M_{\hi}$ the visible \hi~ mass of the cloud.
\end{table*}

The total area covered by the 22 fields observed with the 100-m telescope is
$\sim$204 deg$^2$. Due to the diffuse nature of the clouds it was not possible
to use automated detection algorithms, so the selection was done
manually. Application of the criteria discussed in
Sect. \ref{sec_2a} yielded 25 objects identified as \hi~ halo
clouds with line-of-sight velocities \VLSR~ close to the emission of the
underlying disk. Fig. \ref{12673fig1} shows a few examples of the observed line
spectra. The lines are narrow but superposed on extended wings of diffuse
Galactic \hi~emission. For the 100-m telescope, stray radiation effects have been taken
into account; we are therefore confident that these observed  components
are not caused by instrumental effects. 

Table \ref{table1} summarizes our single dish results. From the observed  cloud position $l, b$  and velocity $v_{lsr}$ we derive its Galactocentric
distance $R$ and the height $z$ above the plane \citep{Kalberla2008AA}. The angular diameter $s$ is the
geometrical mean of the major and the minor axis of the cloud. The high
brightness temperature $T_{B}$ of the clouds ensures that the effect of the noise is
minimal. In some cases where the entry is missing in Table \ref{table1}, the
clouds were unresolved and the diameter could not be constrained due
insufficient beam-by-beam measurements and because of the blending with the extended wings.
The spatial diameter is calculated from  $s$ making use of the known
distance. As mentioned in Sec. \ref{sec_2b} this is one
major source of uncertainties. Assuming optically thin gas, to obtain the line
width \FWHM~ and the column density $N_{HI}$ we tried to fit a one or two component Gaussian to the
average spectrum of the cloud. In some clouds such as e.g. 116.2+23.6,116.7+23.0 this
was successful. In  clouds like e.g. 113.0-12.4 and 113.3-14.1, where the
extended wing was stronger, this was not possible. In those cases we had to first model
the extended wing, then subtract it and  estimate the \FWHM~ and the column
density of the cloud. The observed line width \FWHM~ defines the upper limit
for the kinetic temperature $T_{kin}$ due to the effect of turbulence. For
very cold clouds, assuming a spin temperature $T_{s}=80$ K, we expect the
derived temperature not to be strongly biased, while for the broader lines the
bias is larger.  To determine the average volume density $\langle n \rangle$, we
assumed a cylindrical shape: $\langle n \rangle=N_{HI}/d$. Deviations from this shape and the uncertainties in the
distances determination are the major sources of biases in this estimate. 
The pressure was estimated from $P= \langle n \rangle \cdot T_{kin}$. This includes the
thermal pressure as well as a turbulence component. Finally from the column density $N_{HI}$ and
size of the cloud $D$ we estimated its visible mass using $M=N_{HI}*D^{2}$.\\

\begin{figure*}
\centerline{
\includegraphics[scale=0.4]{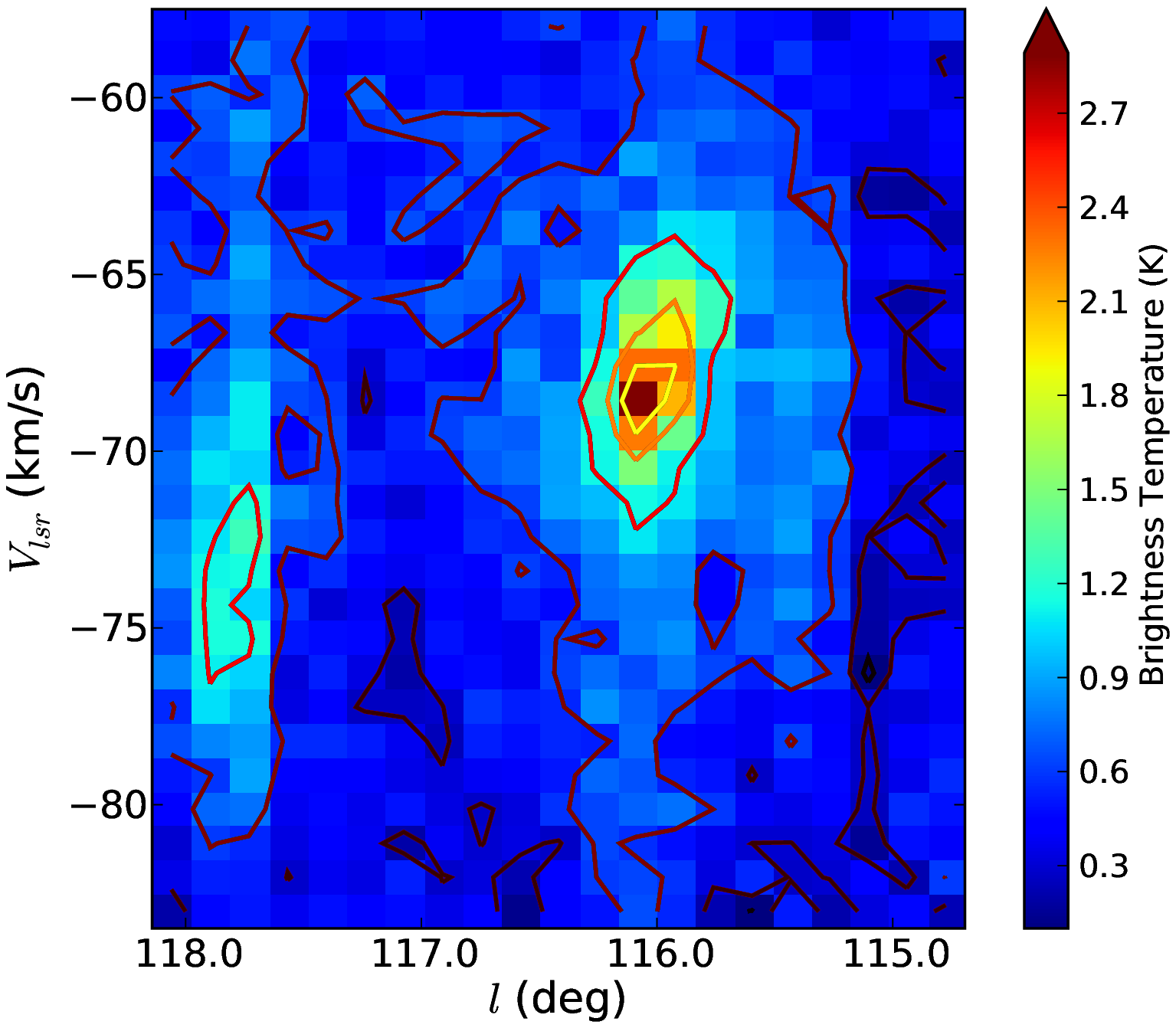}
\includegraphics[scale=0.4]{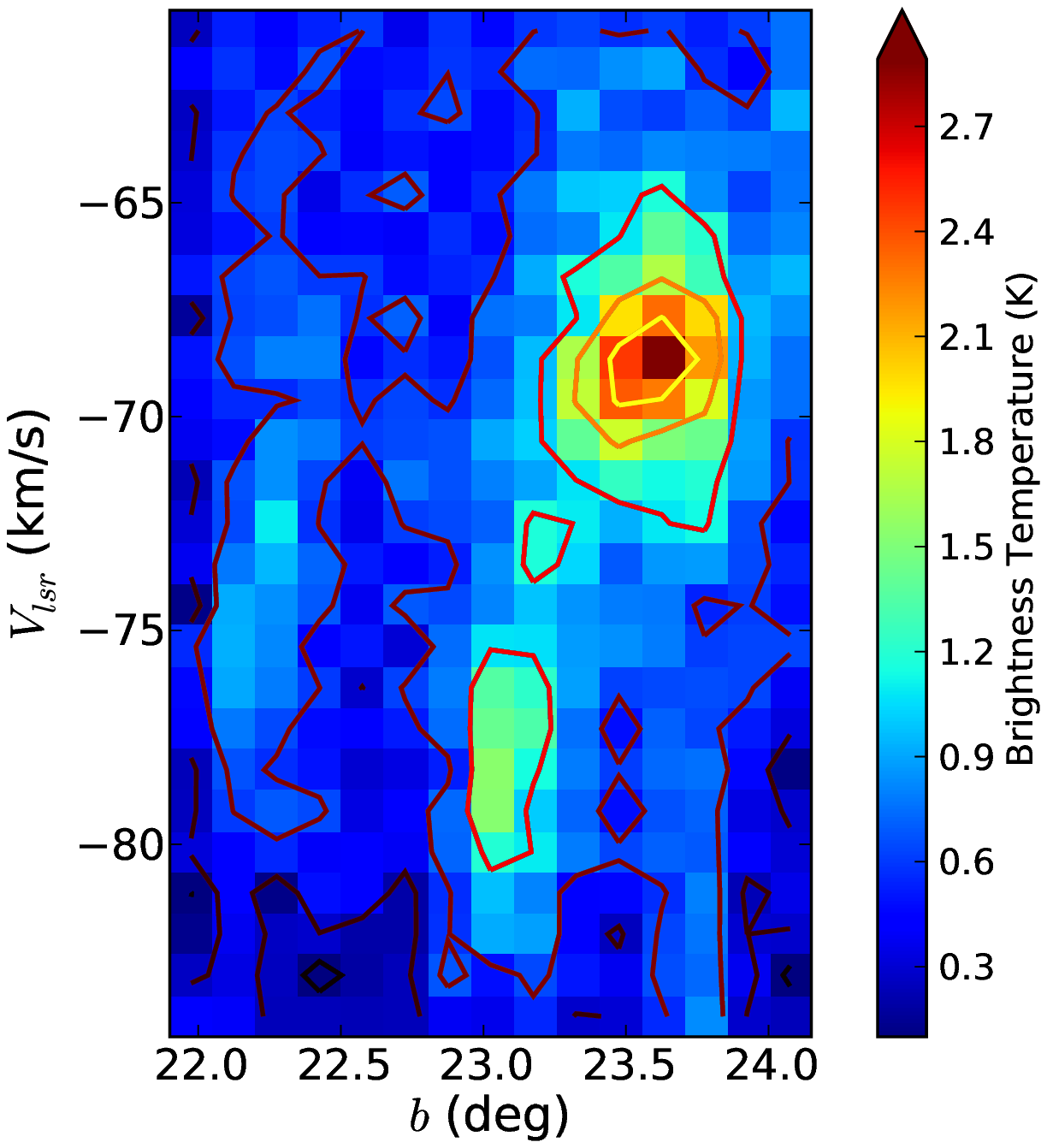}}
\caption{Left: a) A longitude-velocity \hi~ brightness temperature map of the cloud
  116.20+23.55 taken by the Effelsberg 100-m telescope. The \rms~ is 0.1K. The
  color-bar displays the transfer function. The contours are at 0.1, 0.3, 0.6,
  0.9, 1, and 2 K. Right:b) A latitude-velocity \hi~ brightness temperature map of the  cloud 116.20+23.55 taken by the Effelsberg 100-m telescope. The \rms~ is 0.1K. The color-bar displays the transfer function. The contours are at the  level of 0.1K 0.3K 0.6K 0.9K 1K 2K.}
\label{12673fig2}
\end{figure*}

The Gaussian deconvolution of the clouds showed that most of them can
     best be fitted  with a two-component Gaussian. This corresponds to a
     narrow compact component and to a broader more extended component. The
     best two examples which are present in this paper are the cloud
      116.2+23.6 (Fig. \ref{12673fig2}) and 115.0+23.9. All the properties of the
     narrow components are given in Table \ref{table1} except the peak
     temperature, which for  cloud 116.2+23.6 is $T_{C}$=1.8$\pm$0.1~K and for
     the cloud 115.0+23.9 is $T_{C}$=0.73$\pm$0.1~K. We estimated a  peak
     temperature  $T_{W}$=0.71$\pm$0.1~K, column density
     $N_{W}$=25$\pm$4$\cdot$~\unith~ and a \FWHM=18$\pm$0.5~\kms~ for the broad
     component of the cloud 116.2+23.6. For the cloud 115.0+23.9 the Gaussian
     decomposition gave a peak temperature $T_{W}$=0.8$\pm$0.1~K, a column
     density $N_{W}$=20$\pm$4~\unith~ and \FWHM=15$\pm$1~\kms. Due to the
     presence of the Galactic wings in the emission profiles, the size of the broad component is
     uncertain. Assuming that it has a similar dimension as  the narrow part,
     we  have a volume density of $ \langle n \rangle $=0.16$\pm$0.07 $\mathrm{cm^{-3}}$~ 
and   pressure $P$=1151$\pm$507 $K\cdot\mathrm{cm^{-3}}$ for 116.2+23.6 and
volume density $ \langle n \rangle $=0.11$\pm$0.04 and $P$=535$\pm$211 $K\cdot\mathrm{cm^{-3}}$ for 115.0+23.9.\\

According to \citet{Rohlfs2004}, assuming virialization, the line widths \FWHM~ of the \hi~ clouds can
be used to estimate their virial masses. For the \hi~ clouds in
Table \ref{table1} these calculations give typical virial masses
$M_{vir}\sim10^{4}$ \msun, which are more than two orders of
magnitude larger than the visible \hi~ masses listed there.
This comparison indicates that the \hi~ halo clouds observed with the
Effelsberg telescope cannot be self-gravitating objects. Therefore an
external confinement is needed for the clouds not to disperse. Assuming the
presence of a hot halo according to \citet{Pietz1998} and \citet{Kalberla2007DM}, the counterpart
could be provided by the envelope and the surrounding hot halo medium. But in the cloud 116.2+23.6 its pressure $P$=14$\pm$8
 K$\cdot\mathrm{cm^{-3}}$ is small in comparison to the envelope
pressure of $P$=1151$\pm$507 K$\cdot\mathrm{cm^{-3}}$. The same is
true for the cloud 115.00+23.9, where the pressure for the cloud is
$P$=19$\pm$7~K$\cdot\mathrm{cm^{-3}}$~ in comparison to the pressure
of the envelope of $P$=535$\pm$211 K$\cdot\mathrm{cm^{-3}}$. In addition, a comparison of
the \hi~ cloud pressures $P=\langle n \rangle \cdot T$ with theoretical
estimates for warm component pressures from \citet{Wolfire2003} at the given
Galactocentric radius range $ 10 \la R \la 15$ kpc indicate a similar trend
for the rest of the clouds of the sample. Before reaching a firm conclusion
that the \hi~ halo clouds may be transient objects, one can conclude that the
clouds are unresolved.  This implies that estimates for the sizes are upper
limits only and higher resolution observations are needed.

\subsection{Synthesis observations}
\label{sec_4b}

 Two of the \hi~ halo cloud positions from the Effelsberg sample were observed
at high resolution to get a better constraint of the properties of the
clouds. The first cloud 116.2+23.6 was observed with WSRT at a position (J2000)
$\alpha=$20\lfh4\lfm22\lfs, 82\lfd56\lfarcm36\lfarcs~ and has a line-of-sight
velocity \VLSR=-68 \kms, which corresponds to a distance of $d=7$ kpc ($R=13$ kpc,
$z=2.5$ kpc). It is cold with a line width \FWHM=3 \kms~ and a column density
of \NH=$11\pm2$ \unith. The second \hi~ halo cloud, 115.0+23.9,  was observed with
VLA, at an offset position  (J2000) (\ra,\dec) =20\lfh 29\lfm, 82\lfd08\lfarcm~
and has a line-of-sight velocity \VLSR=-84.50 \kms, which corresponds to a
distance of $d$=10 kpc ($R=15$ kpc, $z=4$ kpc). This cloud shows also a narrow line width
\FWHM=3 \kms~ and has a column density \NH=$15\pm3$ \unith.

\begin{figure*}
\centerline{
\includegraphics[clip,trim=0 0 70 50,scale=0.4]{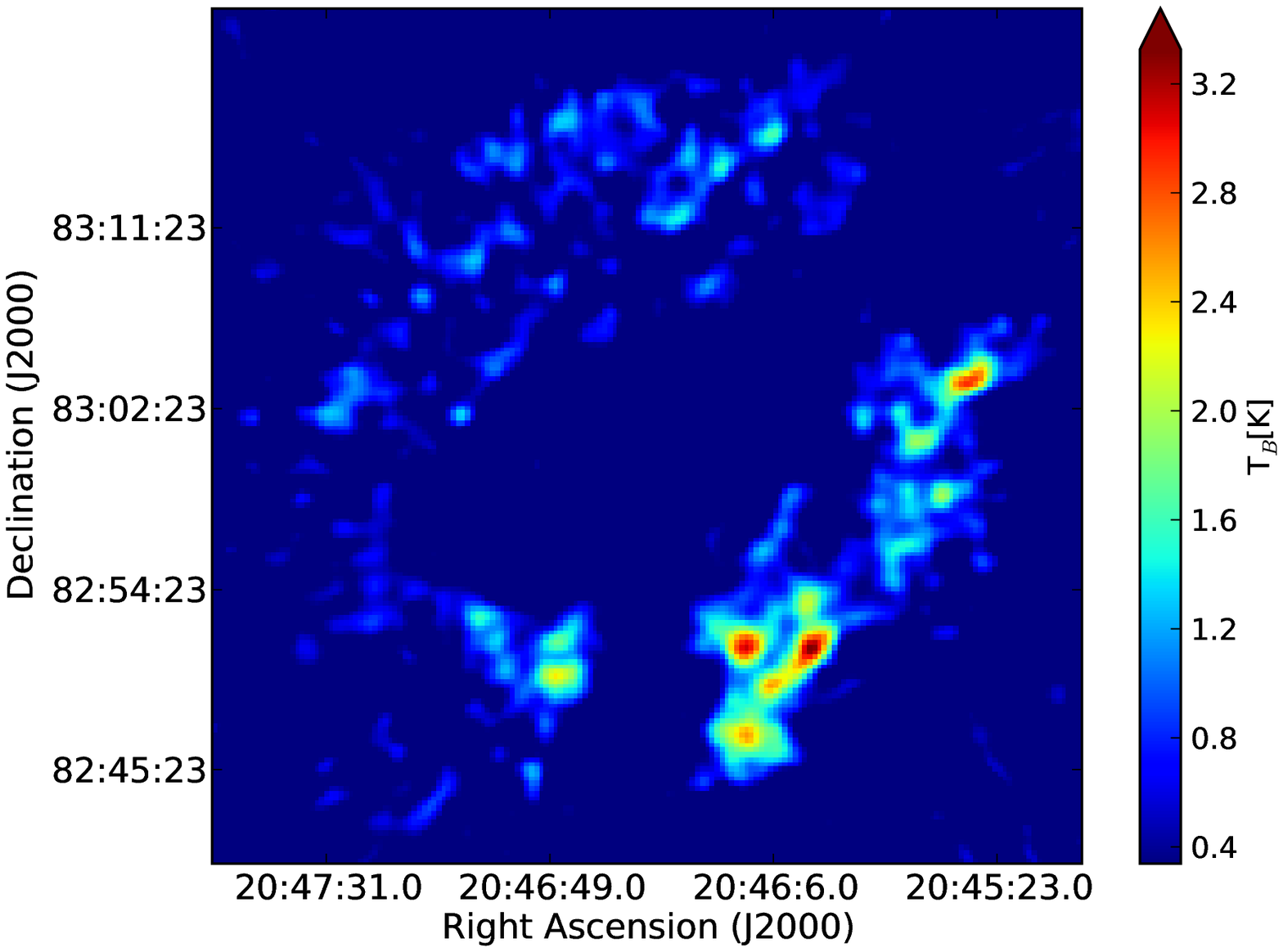}
\includegraphics[clip,trim=0 0 70 50,scale=0.4]{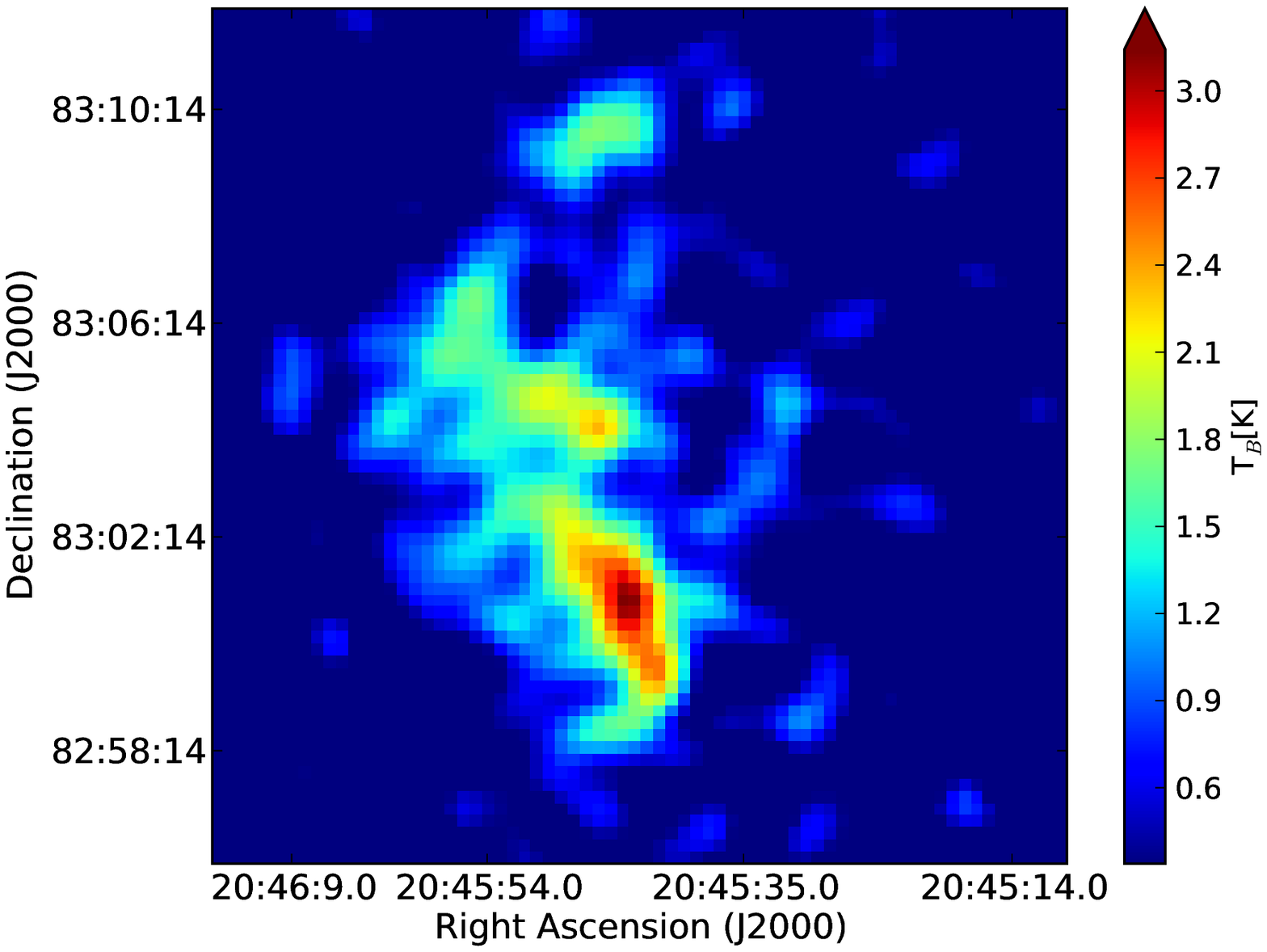}
}
\caption{a)  A RA-DEC map of \hi~  emission observed with the WSRT telescope
  centered at  \ra=20\fh51\fm00\fs, \dec=83\fd01\fm52\fs, \VLSR=-68.72
  \kms. \hi~ emission below the 1-\rms~ of =0.4K has been blanked out.  b) An RA-DEC map of \hi~  emission of the cloud 116.2+23.6 observed with the WSRT telescope centered at  \ra=20\fh43\fm48\fs, \dec=83\fd04\fm44\fs, \VLSR=-84.18 \kms. \hi~ emission below the 1-\rms~ of 0.4K has been blanked out.}
   \label{12673fig3}
 \end{figure*}

\begin{figure}
\includegraphics[clip,trim=10 0 90 40,scale=0.4]{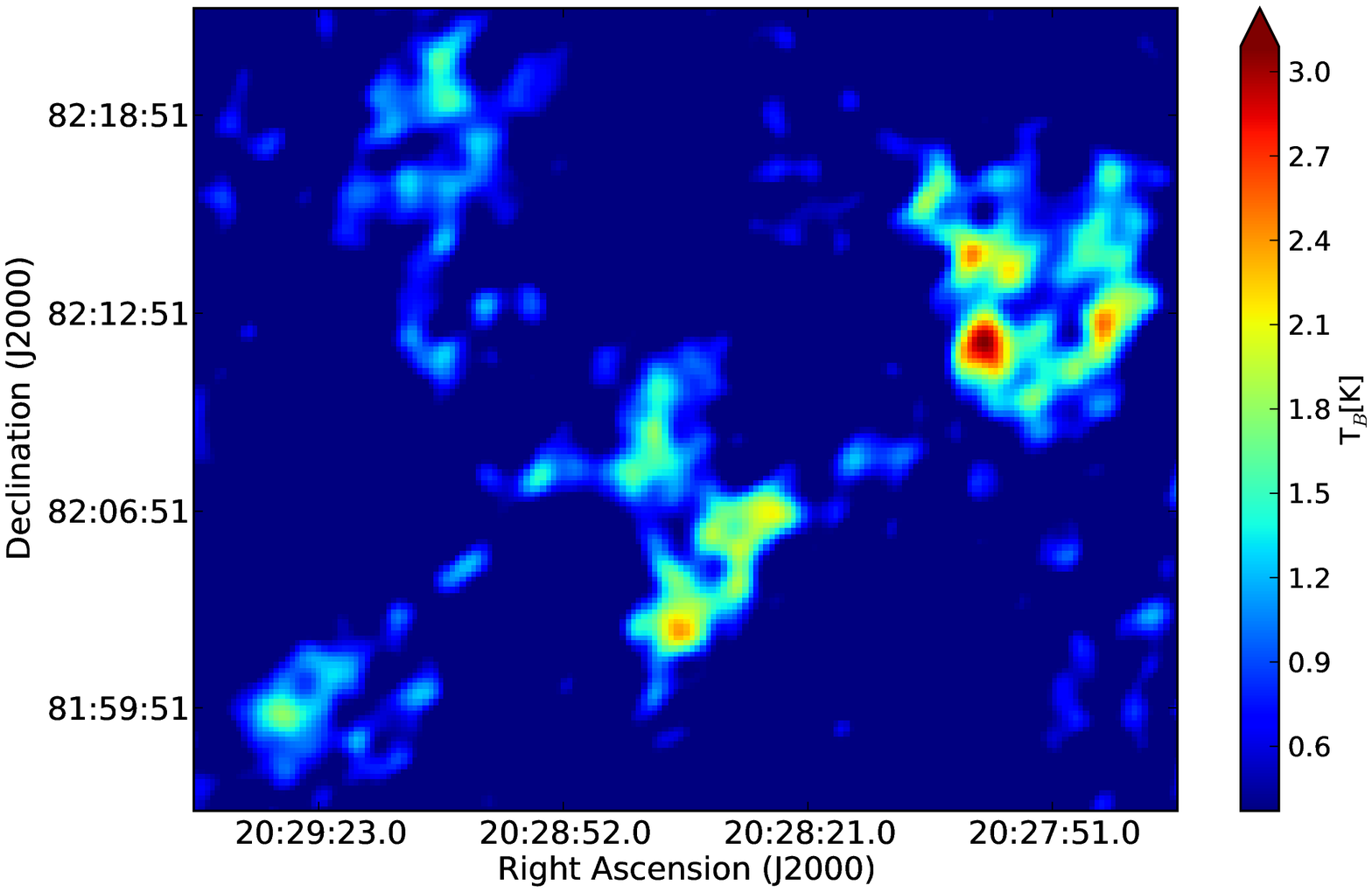}
\caption{An RA-DEC map of \hi~ emission of the cloud 115.0+23.9 observed with VLA array centered at
  \ra=20\fh29\fm00\fs, \dec=82\fd08\fm00\fs, \VLSR=-84.50 \kms. \hi~ emission
  below the 1-\rms~ of 0.4K has been blanked out.}
   \label{12673fig4}
 \end{figure}

As seen in Fig. \ref{12673fig3}a a collection of compact \hi~ objects was  found in the WSRT observations at a \VLSR=-68 \kms. All cores with an angular radius s$\geq60$\arcsec~ are significant at
 a signal-to-noise level of 5 or more. Some unexpected cores were found at a
 line-of-sight velocity of \VLSR=-85 \kms~(Fig.\ref{12673fig3}~b). These are
 associated with an \hi~ emission that is barely significant in the
 Effelsberg spectra. Similarly, at the position (J2000)
 (\ra,\dec)=20\lfh29\lfm, 82\lfd08\lfarcm, observed with the VLA we
 find a conglomeration of at least 8 cores at a velocity of \VLSR=-84 \kms~
 which are associated with the \hi~ cloud 115.0+23.9. In Fig. \ref{12673fig4} 
a channel map of the VLA  observations is given.

 In Table \ref{table2} median values for the observed quantities
 and derived physical parameters respectively are given for the \hi~ cores
 detected with the WSRT and VLA telescopes. To determine the angular sizes 
  and their column densities we produced separated 0th moment maps, and we
  fitted  a 2D Gaussian. The fitted  full width half maximum is the angular size
 $s$ of the cores. From the integrated flux we estimated the column
  density. The line width is determined by inspection of the line since it was
  not possible to fit the spectra. Assuming a spherical shape, the average
  volume density is estimated. Finally for the pressure we used a similar method
 as the one described in Sec. \ref{sec_4a}. The \hi~ cores show very narrow lines
 with a median of \FWHM=3.3 \kms~ and \FWHM=4.3 \kms~ for WSRT and VLA
 respectively, implying cold gas. The \hi~ cores are also very compact with a
 median angular size $s$=78\arcsec~ for WSRT and $s$=88\arcsec~ for VLA,
 implying a median linear diameter of $D$=2.6 pc and $D$=5 pc respectively. For
 the measured column densities which are given in Table \ref{table2}, the
 above diameters imply that the cores are a lot denser than what was estimated
 from the Effelsberg observations. For the \hi~ cores detected with the WSRT, the
 median volume density is $\langle n \rangle$=2.3$\pm$0.8 $\mathrm{cm^{-3}}$, while for the
 \hi~ cores detected with the VLA the median is $\langle n \rangle=3.6\pm1.1 \mathrm{cm^{-3}}$. 
A comparison of the total mass observed  in the \hi~ cores with the total visible \hi~ mass of the ``parent''
 Effelsberg cloud shows that they carry a large fraction of the mass. In the case
 of the WSRT cores they cover an area of 15arc-min$^{2}$ with a total mass of
 33\msun, while the VLA cores cover an area of 14.55 arcmin$^{2}$ and have a
 total visible \hi~ mass of 107 \msun. The \hi~ cores have very small filling
 factors of 3\%-4\%, but carry a significant amount of the mass, up to 48\% in
 comparison with the Effelsberg observations. The rest of the mass is
 associated most probably with the  warmer \hi~ envelope, which is extended and
 undetectable for an interferometer.

\begin{table*}
\caption{Derived parameters of the \hi~ cores.}
\begin{center}
\begin{tabular}{c|| c c c c c c c}
\hline
          &$N_{HI}$ & $s$  &$D$  & $T_{kin}$&  $\langle n \rangle $ &   $P$    &  $M_{\hi}$  \\
          & \unith & arcmin  & pc &  K       & $\mathrm{cm^{-3}}$ &  $K\cdot\mathrm{cm^{-3}}$ & \msun    \\
\hline
\hline
Range     &12--32  &58--127    &1.9--5.9  &140--350 & 1.3--3.1& 270-980 & 1---24 \\
Median      &16  &  78      &2.6 & 240       &  2.3   & 560    & 4  \\ 
Range     &29--76& 54-124     &2--6 &170--730  & 2.2--5.7  &780--2440&7--32 \\
Median      & 40  &   86     & 5   &   410  & 3.6       & 1500   &16   \\
\hline
\end{tabular}
\label{table2}
\end{center}
Note - The first two rows correspond to  the cloud 116.2+23.6 observed with WSRT. The last two rows  correspond to the cloud 115.0+23.9 observed with VLA. $N_{HI}$  is the observed \hi~
  column density of the cores, $s$~ is the angular size, $D$~ is the spatial
  diameter. $T_{kin}$~ is the kinetic temperature. $\langle n \rangle $~ is
  the volume \hi~ density. $P$ is the pressure of the \hi~ gas. $M_{\hi}$ is
  the visible \hi~ mass of the core. 

\end{table*}

The high resolution interferometer observations enable us to determine
accurately the morphology of the \hi~ clouds observed with Effelsberg. They are resolved
into individual \hi~ cores for which kinematical pressures could be
determined. As seen in Table \ref{table2}, the median pressure of $P=560\pm230$
K$\cdot\mathrm{cm^{-3}}$ for the WSRT sample and the median pressure of
$P=1500\pm490$ K$\cdot\mathrm{cm^{-3}}$ for the VLA sample are both comparable
within the uncertainties with the pressures of the surrounding envelope as
estimated from the Effelsberg observation, which are $P$=1151$\pm507$
K$\cdot\mathrm{cm^{-3}}$, and $P$=535$\pm$211 K$\cdot\mathrm{cm^{-3}}$ for the
cloud observed with WSRT and VLA respectively.

A comparison of the pressures for cores and envelopes with theoretical estimates at
different Galactocentric radii from Table 4 of~\citet{Wolfire2003} shows that
the derived quantities fit well to a two-phase picture with an approximate
pressure equilibrium between the cores and the surrounding warm \hi~
envelope.  The envelopes as detected in the single dish observations (see
Fig. \ref{12673fig2}) may provide the necessary support to stabilize the clouds. 
The envelope with a peak brightness temperature of $\sim$0.8 K is too weak
    to be detectable in our interferometer data. Additional deep observations
    would be needed to clarify the presence and extent of such a component. 
Nevertheless, combining the Effelsberg and WSRT/VLA observations can give us an
insight into the nature of the ISM in the region of these clouds. Taking into
account the presence of the very extended broad galactic wings, the warm
envelope and the small scale cold cores which are unresolved with the
Effelsberg telescope,
we reach  a sort of hierarchical structure of the ISM, similar to the
turbulent flows found by \citet{Audit2005}, which will be discussed later.

Using equation 47 from \citet{McKee1977}~ we can show that the
envelope would protect the \hi~ cores from fast evaporation. In the absence of
a warm envelope, the mass loss for an \hi~ core embedded in a hot plasma
(T$\sim10^6K$) is $\sim 7 \cdot 10^{-2}$\msun$ \mathrm{\cdot Myrs^{-1}}$.
This implies that a core with an average mass of $\sim10$ \msun~ will evaporate in $\sim$
140Myrs. A core embedded in a warm envelope ($T \sim 7000$ K \citet{Wolfire2003})
has a mass loss of one magnitude smaller, on the order of $\sim 6 \cdot 10^{-3}$
\msun$\mathrm{\cdot Myrs^{-1}}$, and the cores will evaporate in $\sim$1.7Gyrs. But because of internal turbulent motions the clouds will
evolve within $\sim 2\cdot 10^{5}$yrs. This is  a very short time in comparison
with the total time of their orbit, which is around 100 Myrs, which is found by simple
ballistic simulation.

\subsection{Comparison between the properties of different samples of  \hi~ 
  clouds}
\label{sec_5}  

To better understand  the physical properties of the \hi~ halo clouds, but also
in order to determine various systematics, it is reasonable to compare our
clouds with different samples of \hi~ clouds detected with other
telescopes. The samples of clouds for a comparison are: a) \citet{Lockman2002},
where a population of clouds in the inner Galaxy ($ R \sim 3.5$kpc) was detected
with GBT. The distance was determined using the terminal velocity of the
sources. b) \citet{Snezana2006}, using the Arecibo 300-m radio telescope,
detected a number \hi~ clouds distinctively separated from the Galactic disk
towards the anti-center direction. Due to high intrinsic uncertainties,
kinematic distances were not used, and the authors opted to use pressure
equilibrium considerations to determine the distance. c) \citet{Ford2008ApJ},
where a large number of \hi~ clouds were detected within the pilot region of
the Galactic All-Sky Survey (GASS) with the Parkes 64-m telescope. The
terminal velocity was used as well to determine the distance, since the clouds are
located in the inner Galaxy. d) \citet{Stil2006}, where in the VLA Galactic
Plane Survey (VGPS) 17 fast moving \hi~ clouds were
detected close to the plane which are possibly associated with the halo gas
phase. 

An inter-comparison of the main results from these authors is given in Table
\ref{table3}. Telescope independent quantities, like \NH~ and \FWHM, are found
to be similar, with an exception of \NH~ from \cite{Stil2006}; the higher
column densities in this sample can be explained by the low altitude of the
sample. The other cloud properties are similar, which may imply that the
different cloud samples in the inner and the outer Galaxy may have a similar
origin.

The telescope-dependent properties of the samples, $s$, $D$, $n$, $M_{\hi}$,
show a different picture. Telescopes with similar resolutions, like Effelsberg,
GBT and Parkes, show similar ranges for $s$, $D$, $\langle n \rangle$,
$M_{\hi}$. Taking into account that the source distances are fairly
    similar, it is quite clear that  the  telescope used in each work may
    introduce a twofold systematic bias: a) Synthesis telescopes work as
    spatial filters, more sensitive to cold compact than to warm extended gas. b) 
   The measured angular diameter $s$ depends on the convolution of the actual
   cloud size $s_{t}$ and the beam size. As such the $s_{t}$ is overestimated,
   leading to an underestimate of the measured volume density $n$. For the same
   cloud, telescopes with increased resolution will either resolve the cloud
   or measure $s$ closer to the  $s_{t}$. Therefore the errors in the
   determination of the parameters depending on the angular size such as $D$, $n$,
   $M_{\hi}$ are smaller, and the measurements are closer to the properties of
   the clouds. This dependence on the beam size obviously explains that the Arecibo
observations (beam resolution $\sim4$ \arcmin)  yield parameters for the clouds
closer to the ones estimated for the  \hi~ cores observed with the WSRT and the VLA telescope.\\

\begin{table*}
\begin{center}
\caption{A comparison between the \hi~ halo cloud samples observed with
  different telescopes.}
\begin{tabular}{c||c c c c c c c c}
\hline
    &  $|z|$  & $d$   &  \NH &  \FWHM & $s$ & $D$  & $\langle n \rangle $ & $M_{\hi}$     \\
 & kpc & kpc & $10^{19}\mathrm{cm^{-2}}$ & \kms & \arcmin & pc & $\mathrm{cm^{-3}}$ & \msun  \\
\hline
\hline

Effelsberg & 0.9--5.4  & 3--11  & 0.4--7.4 & 2--21.1   &   9--27  & 8--70   & 0.07--1.2 & 3--770\\ 
WSRT&          &       & 1.2--3.2 & 2.5--4.0  & 0.9--2.1 & 2--6    & 1.3--3.1 & 1--24 \\
VLA&          &       & 2.9--7.6 & 2.8--5.8  & 0.9--2.1 & 2--6    & 2.2--5.7 & 7--32 \\
\citet{Lockman2002}& 0.6--1.2  &       & 0.7--6.3 & 5.4--26.3 &         &19--35   & 0.1--0.9  &12--290 \\
\citet{Snezana2006}& 0.06--0.9 &0.2--3  & 1--3      & 3--7.6  & 6--12    &0.6--8   &          & 0.03--7\\
\citet{Ford2008ApJ}& 0.3--1.7  &       & 0.2--2.2  &5.8--26.2  &         & 46--100 &          & 120--4850 \\ 
\citet{Stil2006}& 0.01--0.16&4.2--7.7& 8--39     &3.4--16.5  & 1.8--29.4&3.4--36.7&3--24      &9--2500\\ 
\hline                 

\end{tabular}
\label{table3}
\end{center}
Note - The table gives the range of values of the clouds for
  the following properties: $|z|$ which is the absolute height above the plane
  (defined in the inner  part of the disk) in kpc. $d$ the distance in
  kpc. \NH is the column   density of \hi~ gas in $10^{19}\mathrm{cm^{-2}}$. \FWHM~ is the line
  width  in \kms. $s$ is the observed angular size  in \arcmin. $D$ is the
  diameter in pc.  $\langle n \rangle $  is the average volume density  in
  $\mathrm{cm^{-3}}$. $M_{\hi}$  is the visible \hi~ mass in \msun.
\end{table*}

\section{Discussion}
\label{sec_6} 

In Sect. \ref{sec_4b} we found indications for a pressure equilibrium
between the \hi~ cores which constitute the halo clouds and the warmer
envelopes in which they  are possibly embedded. In the Galactic disk, it has
been shown \citep[see][and reference therein]{Wolfire2003} that the two main
phases, the cold medium phase with $T\sim100$K and the warm medium phase (WNM)
with $T\sim10^{4}$K \citep{KULKARNI1987}, can co-exist in pressure and thermal
equilibrium only in a very narrow range of pressures and densities. The
\citet{Wolfire2003} model determines this range for different Galactocentric
distances up to $R=18$kpc. The model takes into account various observational
constraints, e.g. dust and metalicities; for a more detailed discussion see
Chap.2 and 3 of \citet{Wolfire2003}. It is assumed that the main heating for the
neutral medium originates from the dust grains through the FUV of young
stars. The cooling of the cold phase is mainly due to the fine-structure of
the CII line (158$\mathrm{\mu m}$), while the cooling of the warm phase
happens in through the Ly$\alpha$, {\ion{C}{II}} (158$\mathrm{\mu m}$) and {\ion{O}{I}}
(63$\mathrm{\mu m}$), with the electron recombination mechanism also playing an
important role. It is important to note here that the model depends on the
dust-to-gas ratios, the metalicity and the assumed FUV field \citep{Wolfire1995}.

Since we were able to determine the volume densities of the \hi~ gas $\langle
n \rangle $ and
the pressure $P$ for the halo clouds and the cores, it is worth comparing the phase diagrams as estimated by \citet{Wolfire2003} with our 
results. We make the following assumptions:

  \begin{enumerate}
  \item In \citet{Wolfire2003} it is assumed that the total density of the
    hydrogen nucleus is $n_{H}=n_{\hi}+n_{H2}$ where $n_{\hi}$ is the
    spatially averaged volume \hi~ density and $n_{H2}$ is the spatially
    averaged molecular hydrogen density. Up to now no direct CO observation of
    the \hi~ halo clouds has been done, therefore their molecular content is
    unknown and we assume that $n_{H}=n_{\hi}$.
  \item The \hi~ halo clouds detected in this work are located at a height of $ 1
    \la z \la 5$ kpc above the plane. According to Eq. 4 from
    \citet{Wolfire1995}, which gives the height dependence of the FUV field,
    the FUV for $z = 5$ kpc is similar to the FUV field in the plane
    ($z=0$). Therefore, for dust-to-gas ratios and metalicities similar to the
    plane, we can use the phase diagrams from \citet{Wolfire2003} estimated
    for the plane ( z = 0kpc).
  \end{enumerate}

\begin{figure*}
\centerline{
\includegraphics[clip,trim=20 0 45 30,angle=0,scale=0.45]{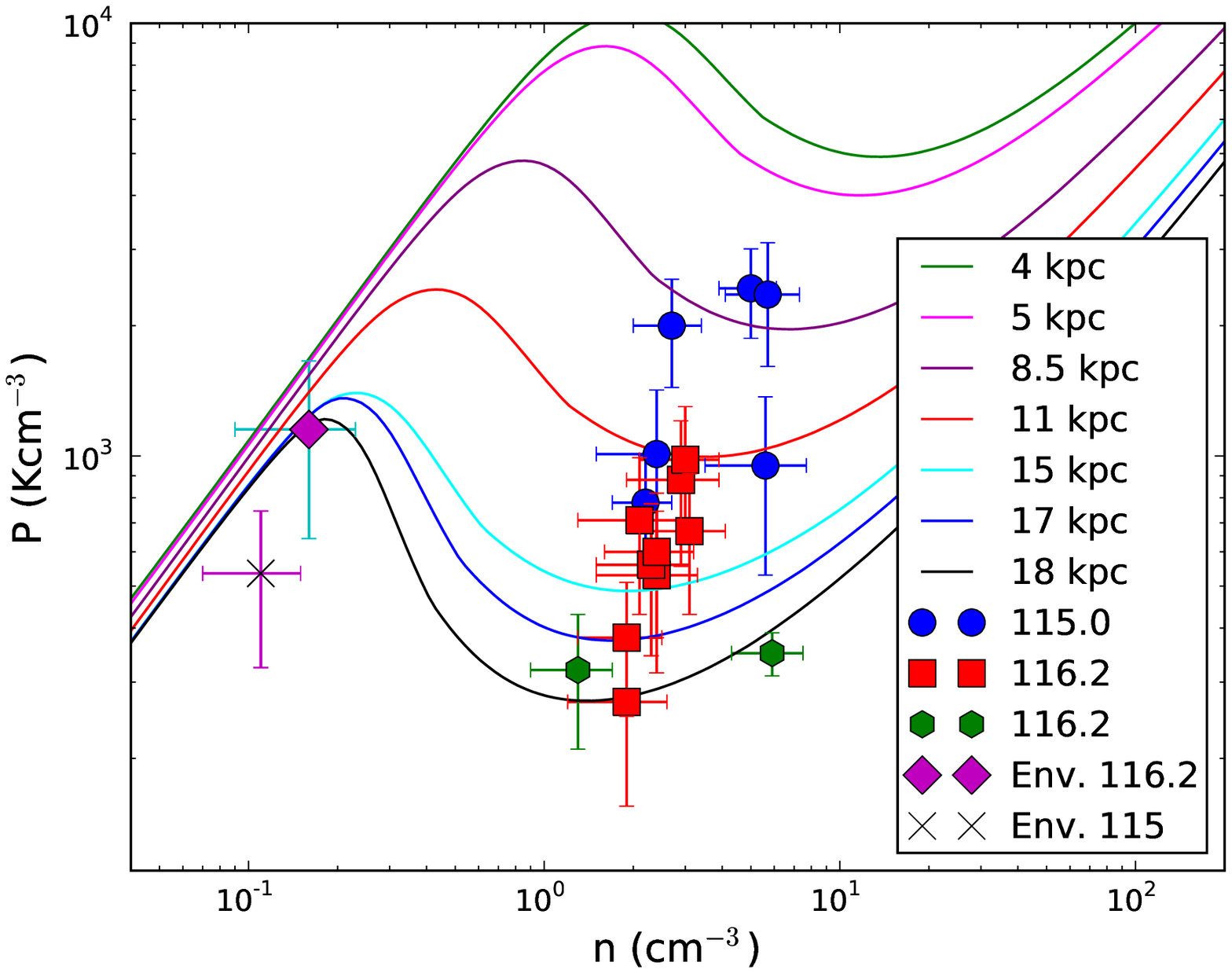}
\includegraphics[clip,trim=20 0 45 30,angle=0,scale=0.45]{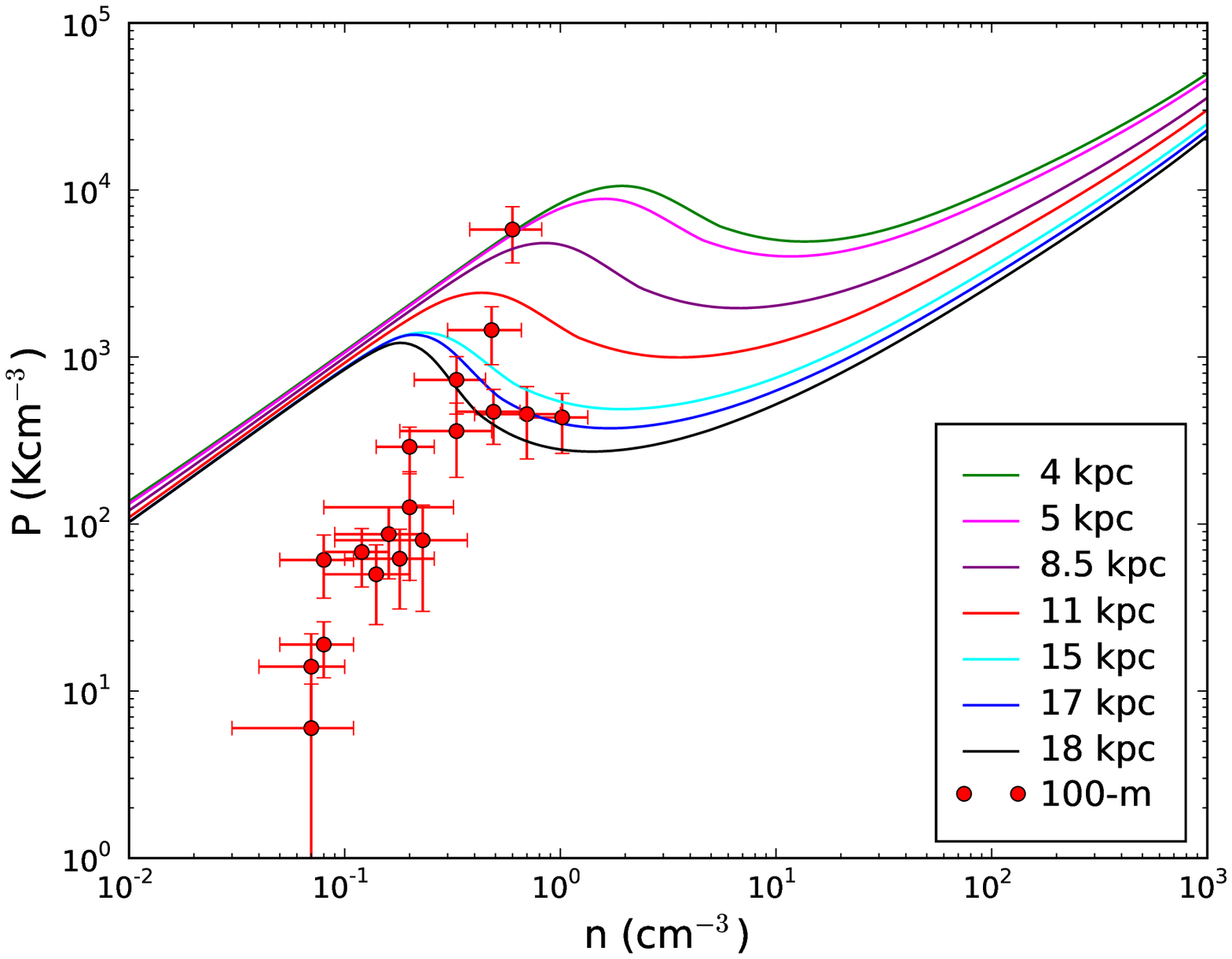}}
\caption{a) Parameters  derived from the WSRT \& VLA samples. The envelope
  pressure is determined by the Effelsberg observations. The curves in both plots apply to
  column densities of the order of 10 $\cdot$\unith~ b) A comparison between the Effelsberg sample of \hi~ clumps and the
  phase diagrams depicting thermal pressure $P$/k vs. hydrogen nucleus density
  $n$ at different Galactocentric radii \citep{Wolfire2003}. }
 \label{12673fig5}
\end{figure*}

 In Fig. \ref{12673fig5}a we compare the properties of the  \hi~cores observed with the VLA/WSRT
 with the theoretical phase diagrams. The envelope estimates in this plot are given from the Effelsberg observations described
 in Sec.\ref{sec_4a} As discussed in  Sec. \ref{sec_4b}, the
 \hi~cores of the VLA have a $R\sim15$ kpc, while
 the WSRT cores have an estimated $R$ from 13 kpc up to 15 kpc. As seen in the
 Fig. \ref{12673fig5}a, at the corresponding $R$ the WSRT cores with \VLSR=-68 \kms~ are
 located barely within the range where thermal equilibrium is possible and
 cold gas can exist in a stable phase.  For the WSRT cores at \VLSR=-84\kms,
 some are located in the thermally unstable region, while others have volume
 densities matching the expectations for the CNM, with smaller pressures than the
 ones expected for the CNM at $R$=15 kpc. For some of the VLA cores in
 Fig. \ref{12673fig5}a we find pressures which are higher than expected for cold gas
 in the equilibrium. 

 The derived pressures, plotted in Fig. \ref{12673fig5}a, may be biased. As described in
Sec.\ref{sec_4a} they were derived from $P =  \langle n \rangle  \cdot T_{kin}$. The uncertainties are
difficult to estimate.  The kinetic temperature is most probably an upper limit, while
the volume density  $\langle n \rangle $ is affected by distance uncertainties, beam
smearing and geometry of the source. Distance uncertainties should be on the order of
20\%. Beam smearing may cause overestimates of the source extension, but
more importantly may be biases caused by the geometry of the source.  If these
clouds have a sheet-like structure \citep{Heiles2003} we may seriously underestimate the volume
density and accordingly the pressure. Both sources show a very
narrow line width. Assuming the typical spin temperature in the range of 40 to
80 K, resulting biases should be within a factor of a few times. Volume densities of the
cores may be more seriously biased. Taking this into account, the true
position of the cores in Fig. \ref{12673fig5}a could possibly be in better agreement with
the cold branch.\\

Regarding the position of the cores in the phase diagram, one other important
detail is that their molecular gas content is unknown. Recent absorption
 experiments report some low column density molecular hydrogen in small
 (0.1pc) and dense clumps in the Milky Way halo with \hi~ column densities $N_H
 > 10^{19}$ \citep{RICHTER2005ASPC}. Such possible biases imply that the cold
 gas in the \hi~ cores can be well in thermal equilibrium with the surrounding
 warmer gas, detected in the single dish observations, which then would play
 the role of a confining envelope.

 The position of the \hi~ cores in the phase diagram is explained in the above
 section, using a simplified static thermal equilibrium hypothesis between the
 two components of the ISM. Non-thermal effects such as turbulence or magnetic
 fields are not taken into account. Recent numerical simulations \citep{Gazol2005,Audit2005} examine the influence of non-thermal factors. As
 a result, the positions of the \hi~ cores in the phase diagram
 (Fig. \ref{12673fig5}a) and their origins can be explained in the context of a
 dynamical equilibrium under the influence of turbulence. \citet{Audit2005}
 examine the influence of turbulence in a converging flow of WNM. Under the
 influence of the turbulent velocity field, warm gas is forced out of thermal
 equilibrium into the unstable regime. In this part of the phase diagram, the
 gas forms cold condensations, which are connected by less dense structure.

 The more turbulent the gas, the larger the fraction of the gas which is
 driven in the intermediate unstable region of the phase diagram  (see
 for comparison in Figs. 3 and 7 of \citet{Audit2005}). A very turbulent flow, also
 part of their simulation, generates more complex structures, and in addition
 cold structures are significantly less dense, even at intermediate
 densities around $\langle n \rangle \sim 5$cm$^{-3}$. Interestingly, this is
 quite similar to the volume densities we observe in the \hi~ cores (Table
 \ref{table2}).  A comparison of the density field and the interferometry maps
 (Figs. \ref{12673fig3}--\ref{12673fig4}) shows a comparable morphology. 

 Similar numerical experiments of \citet{Gazol2005} study the behavior of a
 bistable gas flow under the influence of a turbulent velocity field and model
 gas that is driven into the unstable region. It is shown that as either the
 effective Mach number M or the driving scale increases, a departure of the
 gas from thermal equilibrium is observed, approaching an adiabatic
 behavior. What is interesting in comparison with the phase diagrams presented
 in this work is that a population of under-pressured zones is generated in
 the diffuse gas, while in the dense gas over-pressured zones are
 created. Although, as mentioned before, the simulations do not represent an
 accurate model of the ISM, this trend is probably seen in the phase diagrams
 of Fig. \ref{12673fig5}a. All in all, it seems that the positions of the \hi~ cores
 in the phase diagram agree with the predictions above, implying that
 turbulence  strongly affects the state of the halo ISM. The \hi~ halo clouds
 are probably transient filamentary features, cold unstable gas which is
 continuously condensing out of the WNM, a process that is caused by
 turbulence and is stabilized by it. This result fits well to the finding that
 the extra-planar gas in general is strongly affected and supported by
 turbulence \citep{Kalberla1998}.\\

 A comparison of Fig \ref{12673fig5}a, derived from interferometer data, with
     Fig. \ref{12673fig5}b, using
single dish data only, shows huge differences. Parsec sized \hi~ clumps are
unresolved by single dish telescopes, estimates for pressures and densities
appear seriously biased in this case. Interferometers on the other hand are
insensitive for the extended envelopes which appear to surround cold HI cores.
The ideal telescope should resolve both, the extended features as well as
compact cores. Within a few years the Australian Square Kilometer Array Pathfinder
(ASKAP) may be able to satisfy both demands,
providing a large number of sources in intermediate and high latitudes for a
comparison with theoretical models of the multi phase ISM.\\

\section{Conclusion}
\label{sec_7}

We discussed a population of \hi~ clouds residing in the lower
halo of the Milky Way, co-rotating with the Galactic disk. The sample was
observed with the 100-m Effelsberg telescope. Search criteria were angular
sizes $s$, the brightness temperatures \TB and line width \FWHM~ which are
considered to be typical for halo clumps. The sample includes \hi~ clumps with
the following properties:

\begin{itemize}

\item they reside in the outer galaxy with  Galactocentric radii $R$
  $ 10 < R < 15 $ kpc.
\item they belong to the lower halo ($ 0.9 < z < 5.4$ kpc).
 
\item the gas is cold, with a median \TKIN $ \sim 600$ K and a line width
  \FWHM=5.3\kms.

\item the sample shows a prominent two-component structure. Cold \hi~ cores
  are surrounded by an extended component with broad line emission.
\end{itemize}

Two of the most prominent \hi~ clouds were observed using synthesis arrays,
the WSRT and the VLA. These high-resolution observations resolve the clouds
into a conglomeration of arc-minute sized \hi~ cores. These cores are embedded
in a more diffuse medium which is detectable only with single dish
telescopes. The cores contain a significant fraction of the \hi~ mass and tend
to be in pressure equilibrium with the surrounding envelopes.  Taking into
account the influence of turbulence onto  the line widths \FWHM, the median
line width values of 3.3 \kms and 4.3 \kms observed at the cores implies that
the \hi~ gas is very cold.

Estimating densities and pressures for clumps and surrounding envelopes, we
find some scattering but also a reasonable agreement with models which predict 
pressure equilibrium and a multi-phase structure caused by thermal instabilities
\citep{Wolfire2003}. The clumps tend to populate unstable regions in the phase
diagrams, in agreement with recent predictions of turbulence driven
instabilities \citep{Audit2005,Gazol2005}.

Comparing samples observed with big single dish telescopes e.g. GBT
\citep{Lockman2002}, Effelsberg (this work), and Parkes
\citep{Ford2008ApJ}, we find similar column densities, peak
temperatures, line widths and masses. Our interferometer observations imply
that some of the derived parameters may be heavily biased if the small scale
structure observed by us may be considered as typical for \hi~ halo clumps.

\begin{acknowledgements}
  This publication is based on observations with the 100-m telescope of the
  MPIfR (Max-Planck-Institut f\"ur Radioastronomie) at Effelsberg. For the
  introduction we made use of comments from W. B. Burton concerning early
  observations of the halo gas phase that we received as a response to a
  previous publication. The project was supported by the \emph{Deut\-sche
    For\-schungs\-ge\-mein\-schaft, DFG\/} project number KA1265/5-2. Leonidas
  Dedes would like to thank Karl  Menten for his support and Endrik Kruegel
  for his useful corrections. Finally the authors would like to thank the
  anomymous referee for the his helpful comments.

\end{acknowledgements}

\bibliography{12673bb.bib}
\bibliographystyle{aa}

\end{document}